# Lagged coherence: explicit and testable definition

Roberto D. Pascual-Marqui[1], Kieko Kochi[1], Toshihiko Kinoshita[2]

1: The KEY Institute for Brain-Mind Research; Department of Psychiatry, Psychotherapy, and Psychosomatics; University of Zurich, Switzerland
2: Department of Neuropsychiatry, Kansai Medical University, Osaka, Japan

**Corresponding author: RD Pascual-Marqui**
robertod.pascual-marqui@uzh.ch ; https://www.uzh.ch/keyinst
https://scholar.google.com/citations?user=DDqjOkUAAAAJ

## 1. Abstract

Measures of association between cortical regions based on activity signals provide useful information for studying brain functional connectivity. Difficulties occur with signals of electric neuronal activity, where an observed signal is a mixture, i.e. an instantaneous weighted average of the true, unobserved signals from all regions, due to volume conduction and low spatial resolution. This is why measures of lagged association are of interest, since at least theoretically, "lagged association" is of physiological origin. In contrast, the actual physiological instantaneous zero-lag association is masked and confounded by the mixing artifact. A minimum requirement for a measure of lagged association is that it must not tend to zero with an increase of strength of true instantaneous physiological association. Such biased measures cannot tell apart if a change in its value is due to a change in lagged or a change in instantaneous association. An explicit testable definition for frequency domain lagged connectivity between two multivariate time series is proposed. It is endowed with two important properties: it is invariant to non-singular linear transformations of each vector time series separately, and it is invariant to instantaneous association. As a first sanity check: in the case of two univariate time series, the new definition leads back to the bivariate lagged coherence of 2007 (eqs 25 and 26 in https://doi.org/10.48550/arXiv.0706.1776). As a second stronger sanity check: in the case of a univariate and multivariate vector time series, the new measure presented here leads back to the original multivariate lagged coherence of 2007 (eq 31 in https://doi.org/10.48550/arXiv.0711.1455), which again trivially includes the bivariate case.

## 2. Introduction: general

Consider a collection of signals of electric neuronal activity from different cortical regions in the brain. These can correspond to invasive intracranial recordings, or to the estimated current density from non-invasive EEG/MEG measurements using an electromagnetic tomography method such as those reviewed and compared in Pascual-Marqui et al (2018a).

Such signals can be used for measuring association (i.e. functional connectivity) between cortical regions. This is based on the premise that two regions are functionally connected if their signals are similar (Worsley et al 2005). "Similarity" is quantified with appropriate measures of association.

Invasive recordings can represent localized activity if short-range local bipolar potential differences are used. This is the case of, e.g., ECoG recordings from electrode arrays, when analyses are based on potential differences between very closely located contacts, and not on the raw signals, as demonstrated in Trongnetrpunya et al (2016). This type of data is not commonly available due to its very invasive character.

On the other hand, easily available non-invasive estimated signals are strongly affected by low spatial resolution and volume conduction, meaning that at a particular cortical location, the estimator consists of a weighted linear combination of the cortical activity across the cortex. A desirable activity estimator should have low localization error, with maximum weight at the target location.

Measures of association based on non-invasive signals will seem to be strongly instantaneously similar, mostly due to the mixing artifact. This makes that the actual physiological instantaneous zero-lag association is masked and confounded by the mixing artifact.

This is why measures of lagged association are of interest, since at least theoretically, "lagged association" is of physiological origin. For a measure of lagged association to be of any use, it is minimally required that it must not tend to zero with an increase of strength of instantaneous association. If this requirement is not met, then the lagged association estimator is biased, and cannot tell apart if a change in its value is due to a change in lagged or a change in instantaneous connectivity. When present, this type of bias falsifies the statement that the lagged estimator is not affected by instantaneous association.

It is the aim of this work to present an explicit testable definition for frequency domain lagged connectivity between two multivariate time series. It is endowed with two important properties: it is invariant to non-singular linear transformations of each vector time series separately, and it is invariant to instantaneous association. As a sanity check, in the case of two univariate time series, the new definition leads back to the bivariate lagged coherence in Pascual-Marqui 2007a (see equations 25 and 26 therein).

### 3. Introduction: technical

Basic general theory and methods for the analysis of linear relations between two vector-valued time series can be found in Chapter 8 of Brillinger (2001).

Let $\mathbf{X}_e(t) \in \mathbb{R}^{p \times 1}$ and $\mathbf{Y}_e(t) \in \mathbb{R}^{q \times 1}$ denote two vector-valued time series with discrete time sample $t = 0 \ldots (N_T - 1)$, for epoch (time window) $e = 1 \ldots N_E$. The complex-valued discrete Fourier transforms are denoted $\mathbf{X}_e(\omega) \in \mathbb{C}^{p \times 1}$ and $\mathbf{Y}_e(\omega) \in \mathbb{C}^{q \times 1}$, and defined as:

Eq. 1 $$\mathbf{X}_e(\omega) = \sum_{t=0}^{N_T - 1} \mathbf{X}_e(t) \exp\left(-\iota \frac{2\pi \omega t}{N_T}\right) \quad , \quad \mathbf{Y}_e(\omega) = \sum_{t=0}^{N_T - 1} \mathbf{Y}_e(t) \exp\left(-\iota \frac{2\pi \omega t}{N_T}\right)$$

with discrete frequency $\omega = 0 \ldots (N_T / 2)$ corresponding to cycles per epoch, where $\iota = \sqrt{-1}$.

The cross-spectral covariance estimators for vectors with zero mean are:

Eq. 2 $$\left\{ \begin{array}{l} \mathbf{S}_{xx}(\omega) = \frac{1}{N_E} \sum_{e=1}^{N_E} \mathbf{X}_e(\omega) \mathbf{X}_e^*(\omega) \quad , \quad \mathbf{S}_{yy}(\omega) = \frac{1}{N_E} \sum_{e=1}^{N_E} \mathbf{Y}_e(\omega) \mathbf{Y}_e^*(\omega) \\ \mathbf{S}_{xy}(\omega) = \frac{1}{N_E} \sum_{e=1}^{N_E} \mathbf{X}_e(\omega) \mathbf{Y}_e^*(\omega) \quad , \quad \mathbf{S}_{yx}(\omega) = \frac{1}{N_E} \sum_{e=1}^{N_E} \mathbf{Y}_e(\omega) \mathbf{X}_e^*(\omega) \end{array} \right\}$$

where the superscript "*" denotes complex conjugate and transposed.

Consider the linear relation:

**Eq. 3** $\quad \mathbf{Y}(\omega) = \mathbf{A}_1(\omega)\mathbf{X}(\omega) + \varepsilon(\omega)$

with regression coefficient matrix $\mathbf{A}_1(\omega) \in \mathbb{C}^{q \times p}$, where $\varepsilon(\omega) \in \mathbb{C}^{q \times 1}$ is noise, independent of **X**, and with covariance matrix $\mathbf{S}_{\varepsilon\varepsilon}(\omega) \in \mathbb{C}^{q \times q}$. Such a linear relation in the frequency domain is the result of a general autoregression in the time domain (see e.g. Brillinger 2001, page 287, equation 8.1.5 therein).

The noise covariance is:

**Eq. 4** $\quad \mathbf{S}_{\varepsilon\varepsilon}(\omega; \mathbf{A}_1) = \mathbf{S}_{yy}(\omega) + \mathbf{A}_1(\omega)\mathbf{S}_{xx}(\omega)\mathbf{A}_1^*(\omega) - \mathbf{S}_{yx}(\omega)\mathbf{A}_1^*(\omega) - \mathbf{A}_1(\omega)\mathbf{S}_{xy}(\omega)$

which depends parametrically on $\mathbf{A}_1$.

The least squares estimator for $\mathbf{A}_1$ is derived by solving:

**Eq. 5** $\quad \hat{\mathbf{A}}_1(\omega) = \arg\min_{\mathbf{A}_1} tr\left[\mathbf{S}_{\varepsilon\varepsilon}(\omega; \mathbf{A}_1)\right]$

where the "*tr*" operator returns the trace of the argument. This gives:

**Eq. 6** $\quad \hat{\mathbf{A}}_1(\omega) = \mathbf{S}_{yx}(\omega)\mathbf{S}_{xx}^{-1}(\omega)$

Plugging Eq. 6 into Eq. 4 gives:

**Eq. 7** $\quad \mathbf{S}_{\varepsilon\varepsilon}(\omega) = \mathbf{S}_{yy}(\omega) - \mathbf{S}_{yx}(\omega)\mathbf{S}_{xx}^{-1}(\omega)\mathbf{S}_{xy}(\omega)$

The general unconstrained model in Eq. 3 (with Eq. 6 and Eq. 7) allows for combined "instantaneous and lagged" associations from **X** to **Y**.

Pure zero-lag instantaneous association from **X** to **Y** occurs if the regression coefficient matrix is pure real. This is defined by the constrained model:

**Eq. 8** $\quad \mathbf{Y}(\omega) = \mathbf{A}_0(\omega)\mathbf{X}(\omega) + \delta(\omega)$

where $\mathbf{A}_0(\omega) \in \mathbb{R}^{q \times p}$ is a real-valued regression coefficient matrix. In this case, the noise covariance is:

**Eq. 9** $\quad \mathbf{S}_{\delta\delta}(\omega; \mathbf{A}_0) = \mathbf{S}_{yy}(\omega) + \mathbf{A}_0(\omega)\mathbf{S}_{xx}(\omega)\mathbf{A}_0^T(\omega) - \mathbf{S}_{yx}(\omega)\mathbf{A}_0^T(\omega) - \mathbf{A}_0(\omega)\mathbf{S}_{xy}(\omega)$

where the superscript "*T*" denotes transposed. The least squares estimator for real-valued $\mathbf{A}_0$ is derived by solving:

**Eq. 10** $\quad \hat{\mathbf{A}}_0(\omega) = \arg\min_{\mathbf{A}_0} tr\left[\mathbf{S}_{\delta\delta}(\omega; \mathbf{A}_0)\right]$

giving:

**Eq. 11** $\quad \hat{\mathbf{A}}_0(\omega) = \left[\text{Re}\mathbf{S}_{yx}(\omega)\right]\left[\text{Re}\mathbf{S}_{xx}(\omega)\right]^{-1}$

where the operator "Re" returns the real part of the argument. Note that in general:

**Eq. 12** $\quad \hat{\mathbf{A}}_0 \neq \text{Re}\hat{\mathbf{A}}_1 \Leftrightarrow \left[\text{Re}\mathbf{S}_{yx}(\omega)\right]\left[\text{Re}\mathbf{S}_{xx}(\omega)\right]^{-1} \neq \text{Re}\left[\mathbf{S}_{yx}(\omega)\mathbf{S}_{xx}^{-1}(\omega)\right]$

which means that in general, $\left(\text{Re}\hat{\mathbf{A}}_1\right)$ does not minimize $\left(tr\mathbf{S}_{\delta\delta}\right)$ in Eq. 9 and Eq. 10.

Plugging Eq. 11 into Eq. 9 gives:

**Eq. 13** $\quad \begin{cases} \mathbf{S}_{\delta\delta}(\omega) = \mathbf{S}_{yy}(\omega) + \left[\text{Re}\mathbf{S}_{yx}(\omega)\right]\left[\text{Re}\mathbf{S}_{xx}(\omega)\right]^{-1}\mathbf{S}_{xx}(\omega)\left[\text{Re}\mathbf{S}_{xx}(\omega)\right]^{-1}\left[\text{Re}\mathbf{S}_{xy}(\omega)\right] \\ -\mathbf{S}_{yx}(\omega)\left[\text{Re}\mathbf{S}_{xx}(\omega)\right]^{-1}\left[\text{Re}\mathbf{S}_{xy}(\omega)\right] - \left[\text{Re}\mathbf{S}_{yx}(\omega)\right]\left[\text{Re}\mathbf{S}_{xx}(\omega)\right]^{-1}\mathbf{S}_{xy}(\omega) \end{cases}$

## 4. Definition: Lagged coherence (i.e. lagged association, lagged connectivity)

<u>*Definition: Lagged connectivity*</u>: If $\mathbf{S}_{\varepsilon\varepsilon}(\omega) < \mathbf{S}_{\delta\delta}(\omega)$, we say that there is lagged association from **X** to **Y**.

The measure of lagged association from **X** to **Y** is defined as:

**Eq. 14** $\quad lagA_{y \leftarrow x}(\omega) = \ln \dfrac{\det \mathbf{S}_{\delta\delta}(\omega)}{\det \mathbf{S}_{\varepsilon\varepsilon}(\omega)}$

Lagged coherence is defined as:

**Eq. 15** $\quad lagC_{y \leftarrow x}(\omega) = 1 - \dfrac{\det \mathbf{S}_{\varepsilon\varepsilon}(\omega)}{\det \mathbf{S}_{\delta\delta}(\omega)}$

In this definition, "det" denotes the determinant of the matrix argument, with $\mathbf{S}_{\varepsilon\varepsilon}$ from Eq. 7, and with $\mathbf{S}_{\delta\delta}$ from Eq. 13.

Ordering of matrices, such as $\mathbf{S}_{\varepsilon\varepsilon}(\omega) < \mathbf{S}_{\delta\delta}(\omega)$ in the definition, is explained in e.g. page 287, equations 8.2.1 through 8.2.6 in Brillinger (2001).

The general form of the transformation used from Eq. 14 to Eq. 15 was denoted as the coefficient of determination by Pierce (1982).

Another valid measure of lagged association that similarly compares the two residual covariance matrices is:

**Eq. 16** $\quad lagB_{y \leftarrow x}(\omega) = \dfrac{1}{q} tr \left[ \mathbf{S}_{\varepsilon\varepsilon}(\omega) \mathbf{S}_{\delta\delta}^{-1}(\omega) - \mathbf{I} \right]^2$

where **I** is the identity matrix of appropriate dimension. This measure is related to Nagao's trace criterion (Nagao 1973).

## 5. Lagged association for frequency bands

The estimators for the complex-valued covariance matrices given by Eq. 2 correspond to a single discrete frequency. The covariance matrices for a frequency band, defined by a set "Ω" of two or more discrete frequencies, consist of the sum over the frequencies, e.g.:

**Eq. 17** $\quad \mathbf{S}_{xx}(\boldsymbol{\Omega}) = \sum_{\omega \in \boldsymbol{\Omega}} \mathbf{S}_{xx}(\omega)$

and similarly for $\mathbf{S}_{yy}, \mathbf{S}_{xy}, \mathbf{S}_{yx}$.

Plugging these frequency band covariances into Eq. 7 and Eq. 13 give $\mathbf{S}_{\delta\delta}(\boldsymbol{\Omega})$ and $\mathbf{S}_{\varepsilon\varepsilon}(\boldsymbol{\Omega})$, which in turn can be used for defining the measures of lagged association for the band:

Eq. 18
$$\begin{cases} lagA_{y \leftarrow x}(\Omega) = \ln \frac{\det \mathbf{S}_{\delta\delta}(\Omega)}{\det \mathbf{S}_{\varepsilon\varepsilon}(\Omega)} \\ lagC_{y \leftarrow x}(\Omega) = 1 - \frac{\det \mathbf{S}_{\varepsilon\varepsilon}(\Omega)}{\det \mathbf{S}_{\delta\delta}(\Omega)} \\ lagB_{y \leftarrow x}(\Omega) = \frac{1}{q} tr \left[ \mathbf{S}_{\varepsilon\varepsilon}(\Omega) \mathbf{S}_{\delta\delta}^{-1}(\Omega) - \mathbf{I} \right]^2 \end{cases}$$

## 6. Inference and invariance

### 6.A. Inference

Under Gaussianity, the likelihood ratio test for:

Eq. 19 $\quad H_0 : \mathbf{A}_0 \in \mathbb{R}^{q \times p} \quad, \quad in \quad \mathbf{Y}(\omega) = \mathbf{A}_0(\omega)\mathbf{X}(\omega) + \delta(\omega)$

against the alternative:

Eq. 20 $\quad H_1 : \mathbf{A}_1 \in \mathbb{C}^{q \times p} \quad, \quad in \quad \mathbf{Y}(\omega) = \mathbf{A}_1(\omega)\mathbf{X}(\omega) + \varepsilon(\omega)$

is based on the statistic:

Eq. 21 $\quad N_E \ln \frac{\det \mathbf{S}_{\delta\delta}(\omega)}{\det \mathbf{S}_{\varepsilon\varepsilon}(\omega)} \equiv (N_E)\left[ lagA_{y \leftarrow x}(\omega) \right] \sim \chi^2_{(q \times p)}$

which is asymptotically distributed as chi-square with $(q \times p)$ degrees of freedom under the null hypothesis of pure zero-lag instantaneous association.

If the null hypothesis is true, i.e. if the regression coefficient matrix is pure real-valued, then the statistic in Eq. 21 equals zero.

### 6.B. Invariance to non-singular linear transformations

The measures of lagged association Eq. 14 and Eq. 15 are invariant to any real-valued non-singular linear transformation of **X** and of **Y**. This means that for any non-singular matrices $\mathbf{M}_x \in \mathbb{R}^{p \times p}$ and $\mathbf{M}_y \in \mathbb{R}^{q \times q}$, the transformed data $(\mathbf{M}_x \mathbf{X})$ and $(\mathbf{M}_y \mathbf{Y})$ is identically associated as defined in Eq. 14 and Eq. 15. The demonstration is similar to the result in e.g. Anderson (2003; page 385 to 386, equations 24 to 26 therein).

### 6.C. Invariance to instantaneous association

Consider the particular time-domain Granger-causal autoregressive model of order *m*:

Eq. 22 $\quad \mathbf{Y}(t) = \mathbf{B}\mathbf{X}(t) + \sum_{k=1}^{m} \mathbf{D}(k)\mathbf{X}(t-k) + \varepsilon(t)$

with $\mathbf{X} \in \mathbb{R}^{p \times 1}$; $\mathbf{Y}, \varepsilon \in \mathbb{R}^{q \times 1}$; $\mathbf{B}, \mathbf{D} \in \mathbb{R}^{q \times p}$, see e.g. Granger (1969, equation 3.3 therein), Brillinger (2001, page 287, equation 8.1.5 therein), and Geweke (1982, equation 2.10 therein).

In Eq. 22, the regression coefficient matrices $\mathbf{B}$ and $\mathbf{D}$ are real-valued, with $\mathbf{B}$ corresponding to causal zero-lag instantaneous association from $\mathbf{X}$ to $\mathbf{Y}$, and with all $\mathbf{D}(k)$ for $k=1...m$ related to causal lagged association from $\mathbf{X}$ to $\mathbf{Y}$.

The model in Eq. 22 allows complete control and non-ambiguous separation of instantaneous and lagged causal associations.

Its frequency domain form is:

Eq. 23 $\quad \mathbf{Y}(\omega) = \mathbf{B}\mathbf{X}(\omega) + \mathbf{C}(\omega)\mathbf{X}(\omega) + \varepsilon(\omega)$

with complex-valued $\mathbf{C}(\omega) \in \mathbb{C}^{q \times p}$:

Eq. 24 $\quad \mathbf{C}(\omega) = \left[ \sum_{k=1}^{m} \mathbf{D}(k) \exp\left( -\iota \frac{2\pi\omega k}{N_T} \right) \right]$

The model in Eq. 23 is the same as the basic model in Eq. 3, with:

Eq. 25 $\quad \mathbf{A}_1(\omega) \equiv \left[ \mathbf{B} + \mathbf{C}(\omega) \right]$

Eq. 23 is a generative model for $\mathbf{Y}$, with given and fixed parameters $\left[ \mathbf{S}_{\varepsilon\varepsilon}, \mathbf{S}_{xx}, \mathbf{B}, \mathbf{C}(\omega) \right]$. The population covariances related to the dependent variable $\mathbf{Y}$ are:

Eq. 26 $\quad \mathbf{S}_{yy} = (\mathbf{B}+\mathbf{C}) \mathbf{S}_{xx} (\mathbf{B}^T + \mathbf{C}^*) + \mathbf{S}_{\varepsilon\varepsilon}$

Eq. 27 $\quad \mathbf{S}_{yx} = (\mathbf{B}+\mathbf{C}) \mathbf{S}_{xx}$

where the frequency variable $\omega$ has been dropped for simplification.

The measures of lagged association and lagged coherence in Eq. 14 and Eq. 15 are invariant (i.e. not affected by) any instantaneous zero-lag contribution if the constrained noise covariance $\mathbf{S}_{\delta\delta}$ in Eq. 9 is independent the real-valued matrix $\mathbf{B}$ in Eq. 22 and Eq. 23. This is the only requirement for invariance, since $\mathbf{S}_{\varepsilon\varepsilon}$ is given and fixed.

Plugging Eq. 26 and Eq. 27 into Eq. 11 gives:

Eq. 28 $\quad \mathbf{A}_0 = \left[ \operatorname{Re} \mathbf{S}_{yx} \right] \left[ \operatorname{Re} \mathbf{S}_{xx} \right]^{-1} = (\mathbf{B} + \operatorname{Re}\mathbf{C}) - (\operatorname{Im}\mathbf{C})(\operatorname{Im}\mathbf{S}_{xx})(\operatorname{Re}\mathbf{S}_{xx})^{-1}$

where the operator "Im" returns the imaginary part of the argument. And plugging Eq. 26, Eq. 27, and Eq. 28 into Eq. 9 gives (after several steps of matrix algebra manipulations):

Eq. 29 $\quad \mathbf{S}_{\delta\delta} = \mathbf{S}_{\varepsilon\varepsilon} + \mathbf{C}\mathbf{S}_{xx}\mathbf{C}^* + \mathbf{D}\mathbf{S}_{xx}\mathbf{D}^T - \mathbf{C}\mathbf{S}_{xx}\mathbf{D}^T - \mathbf{D}\mathbf{S}_{xx}\mathbf{C}^*$

with:

Eq. 30 $\quad \mathbf{D} = (\operatorname{Re}\mathbf{C}) - (\operatorname{Im}\mathbf{C})(\operatorname{Im}\mathbf{S}_{xx})(\operatorname{Re}\mathbf{S}_{xx})^{-1}$

The result in Eq. 29 proves the invariance: the new measure of lagged association is not affected by instantaneous zero-lag association.

## 7. The bivariate case (*p*=*q*=1)

For the bivariate case, with univariate signal *x* and *y*, Eq. 7 and Eq. 13 give:

Eq. 31 $\quad s_{\varepsilon\varepsilon} = s_{yy} - \dfrac{(\operatorname{Re} s_{xy})^2 + (\operatorname{Im} s_{xy})^2}{s_{xx}}$

**Eq. 32** $\quad s_{\delta\delta} = s_{yy} - \dfrac{(\mathrm{Re}\, s_{xy})^2}{s_{xx}}$

where the frequency variable ω is present but omitted. Plugging Eq. 31 and Eq. 32 into Eq. 14 and Eq. 15 gives:

**Eq. 33** $\quad lagA_{y \leftarrow x}(\omega) = \ln \dfrac{s_{yy} s_{xx} - (\mathrm{Re}\, s_{xy})^2}{s_{yy} s_{xx} - (\mathrm{Re}\, s_{xy})^2 - (\mathrm{Im}\, s_{xy})^2} = \ln \dfrac{1 - (\mathrm{Re}\, c_{xy})^2}{1 - (\mathrm{Re}\, c_{xy})^2 - (\mathrm{Im}\, c_{xy})^2}$

**Eq. 34** $\quad lagC_{y \leftarrow x}(\omega) = \dfrac{(\mathrm{Im}\, s_{xy})^2}{s_{yy} s_{xx} - (\mathrm{Re}\, s_{xy})^2} = \dfrac{(\mathrm{Im}\, c_{xy})^2}{1 - (\mathrm{Re}\, c_{xy})^2}$

where $c_{xy}$ is the complex-valued coherence:

**Eq. 35** $\quad c_{xy} = \dfrac{s_{xy}}{\sqrt{s_{xx} s_{yy}}}$

Eq. 34 is identical to the lagged coherence of Pascual-Marqui (2007a; equation 25 therein), and Pascual-Marqui (2007b; equation 28 therein). Several useful properties of the lagged coherence were later reported in Nolte et al (2020), while Hindriks (2021) presented a very detailed analysis of its statistical properties.

This present study gives a new independent derivation and justification for the classic lagged coherence (2007), based on a testable definition. It is worth emphasizing once more that the bivariate lagged coherence is not affected by instantaneous zero-lag associations. It enjoys another essential property: it is invariant to any form of mixture for the two signals, as was shown in Pascual-Marqui et al (2018b).

From Eq. 21, the likelihood ratio test for "zero lagged association" is based on the statistic:

**Eq. 36** $\quad (N_E)\left[lagA_{y \leftarrow x}\right] = N_E \ln \dfrac{s_{\delta\delta}}{s_{\varepsilon\varepsilon}} \sim \chi_1^2$

An alternative is the F-test:

**Eq. 37** $\quad F_{1, N_E - 3} = (N_E - 3) \dfrac{s_{\delta\delta} - s_{\varepsilon\varepsilon}}{s_{\varepsilon\varepsilon}}$

equivalent to:

**Eq. 38** $\quad F_{1, N_E - 3} = (N_E - 3) \dfrac{(\mathrm{Im}\, c_{xy})^2}{1 - (\mathrm{Re}\, c_{xy})^2 - (\mathrm{Im}\, c_{xy})^2}$

where $(\mathrm{Re}\, c_{xy})$ and $(\mathrm{Im}\, c_{xy})$ are the real and imaginary parts of the complex valued coherence in Eq. 35.

## 8. Lagged association of a univariate time series with a multivariate vector time series

In this case, let $y(\omega) \in \mathbb{C}$ denote the Fourier transform of the univariate time series, and let $\mathbf{X}(\omega) \in \mathbb{C}^{p \times 1}$ with $p > 1$ denote the Fourier transform of the multivariate vector, related by the unconstrained linear model:

**Eq. 39** $\quad y(\omega) = \mathbf{A}_1(\omega) \mathbf{X}(\omega) + \varepsilon(\omega)$

with $\mathbf{A}_1(\omega) \in \mathbb{C}^{1 \times p}$ and $\varepsilon(\omega) \in \mathbb{C}$. The variance of the noise (see Eq. 7) is:

**Eq. 40** $\quad s_{\varepsilon\varepsilon}(\omega) = s_{yy}(\omega) - \mathbf{S}_{yx}(\omega) \mathbf{S}_{xx}^{-1}(\omega) \mathbf{S}_{xy}(\omega)$

with $\mathbf{S}_{yx}(\omega) \in \mathbb{C}^{1 \times p}$.

The corresponding constrained model:

**Eq. 41** $\quad y(\omega) = \mathbf{A}_0(\omega) \mathbf{X}(\omega) + \delta(\omega)$

with real-valued $\mathbf{A}_0(\omega) \in \mathbb{R}^{1 \times p}$ and $\delta(\omega) \in \mathbb{C}$ has noise variance (see Eq. 13) that simplifies to:

**Eq. 42** $\quad s_{\delta\delta}(\omega) = s_{yy}(\omega) - \left[\mathrm{Re}\mathbf{S}_{yx}(\omega)\right]\left[\mathrm{Re}\mathbf{S}_{xx}(\omega)\right]^{-1}\left[\mathrm{Re}\mathbf{S}_{xy}(\omega)\right]$

Plugging Eq. 40 and Eq. 42 into Eq. 14 and Eq. 15 gives, after some simplification:

**Eq. 43** $\quad lagA_{y \leftarrow x}^{1 \leftarrow p}(\omega) = \ln \frac{1 - R_{y \leftarrow x}^2(\omega)}{1 - C_{y \leftarrow x}^2(\omega)}$

**Eq. 44** $\quad lagC_{y \leftarrow x}^{1 \leftarrow p}(\omega) = \frac{C_{y \leftarrow x}^2(\omega) - R_{y \leftarrow x}^2(\omega)}{1 - R_{y \leftarrow x}^2(\omega)}$

where:

**Eq. 45** $\quad C_{y \leftarrow x}^2(\omega) = \left[s_{yy}^{-1}(\omega)\right]\left[\mathbf{S}_{yx}(\omega)\right]\left[\mathbf{S}_{xx}(\omega)\right]^{-1}\left[\mathbf{S}_{xy}(\omega)\right]$

**Eq. 46** $\quad R_{y \leftarrow x}^2(\omega) = \left[s_{yy}^{-1}(\omega)\right]\left[\mathrm{Re}\mathbf{S}_{yx}(\omega)\right]\left[\mathrm{Re}\mathbf{S}_{xx}(\omega)\right]^{-1}\left[\mathrm{Re}\mathbf{S}_{xy}(\omega)\right]$

where Eq. 46, for real-valued variances and covariances, is known as the squared coefficient of multiple correlation (Brillinger 2001, page 289, equation 8.2.18 therein); and Eq. 45 is known as the *complex-valued case* of the squared coefficient of multiple correlation (Brillinger 2001, page 293, equation 8.2.43 therein).

In Pascual-Marqui (2007b), equation 31 therein, a first version of the multivariate lagged coherence was introduced:

**Eq. 47** $\quad \rho_{x \rightleftarrows y}^2(\omega) = 1 - \frac{\det\begin{pmatrix} \mathbf{S}_{xx}(\omega) & \mathbf{S}_{xy}(\omega) \\ \mathbf{S}_{yx}(\omega) & \mathbf{S}_{yy}(\omega) \end{pmatrix} \det\begin{pmatrix} \mathrm{Re}\mathbf{S}_{xx}(\omega) & \mathbf{0} \\ \mathbf{0} & \mathrm{Re}\mathbf{S}_{yy}(\omega) \end{pmatrix}}{\det\begin{pmatrix} \mathbf{S}_{xx}(\omega) & \mathbf{0} \\ \mathbf{0} & \mathbf{S}_{yy}(\omega) \end{pmatrix} \det\begin{pmatrix} \mathrm{Re}\mathbf{S}_{xx}(\omega) & \mathrm{Re}\mathbf{S}_{xy}(\omega) \\ \mathrm{Re}\mathbf{S}_{yx}(\omega) & \mathrm{Re}\mathbf{S}_{yy}(\omega) \end{pmatrix}}$

It is noted that this previous 2007 definition, for the general multivariate case, is different from (not equal to) the new result in Eq. 15 here. However, they are identical in two particular cases: two univariate time series; and a univariate with a multivariate vector time series.

It can be trivially shown that Eq. 47 is identical to Eq. 44, by simply setting a univariate $y(\omega) \in \mathbb{C}$ and a multivariate $\mathbf{X}(\omega) \in \mathbb{C}^{p \times 1}$ with $p > 1$ in Eq. 47.

## 9. Concluding remarks

### 9.A.

A new explicit testable definition for lagged association is given here. It applies to two multivariate time series. It is truly invariant to (i.e. truly not affected by) instantaneous association. The case of two univariate signals leads to the classic lagged coherence from 2007 (Pascual-Marqui 2007a and 2007b). The case of a univariate time series as a function of a multivariate vector time series leads back to the original multivariate lagged coherence (Pascual-Marqui 2007b, equation 31 therein).

### 9.B.

In a forthcoming study, a comparison to other competing methods will be available.

### 9.C.

It is to be noted that these methods can be applied in different important settings. For instance, consider single trial event-related potentials and their estimated cortical current density signals obtained by some electromagnetic tomography. By computing the discrete Hilbert transform for some given frequency band, lagged association can be computed at each discrete time sample over all repeated presentations of the stimulus. Similarly, the same idea can be applied to data from event related synchronization and desynchronization studies (see e.g. Pfurtscheller and Lopes Da Silva 1999).

### 9.D.

**Measures of lagged phase association**: Note that all the new measures presented here can be applied to pure complex-valued phase data. For instance, elements of the complex-valued vector $\mathbf{X}(\omega)$ consist of the discrete Fourier transforms of each signal. By simply dividing each coefficient by its modulus, the new vector is a vector of complex-valued phases. The phase vectors can be plugged into all equations above, giving measures of lagged association between phases. This is commonly known as phase synchronization, phase locking value, phase coherence, etc. The measures of lagged phase synchronization for the bivariate case were already introduced in Pascual-Marqui (2007a), and measures for the multivariate lagged phase synchronization in Pascual-Marqui (2007b).

### 9.E.

The existence of true physiological instantaneous zero-lag association mechanisms is probably linked to ephaptic transmission, see e.g. Zhang et al (2014), Pinotsis and Miller (2023), and Pinotsis et al (2023). Lagged association is likely to be more related to synaptic transmission.

## 11. History/Version info

- (2023-11-24): First original version #1.
- (2023-11-27): Second version #2: Added subsection "*8. Lagged association of a univariate time series with a multivariate vector time series*".
- (2024-01-07): Third version #3: Current version. Eq. 44 now correct without "logarithm".